# Engagement in Foundational Computer Science Courses Through Supplementary Content for Algorithms

Christopher Birster
Georgia Institute of Technology
Educational Technology
Fall 2017
cbirster3@gatech.edu

**Abstract**

Engaging students in teaching foundational Computer Science concepts is vital for the student's continual success in more advanced topics in the field. An idea of a series of Jupyter notebooks was conceived as a way of using Bloom's Taxonomy to reinforce concepts taught in an introductory algorithms class. The idea of the notebook is to keep the student's engaged in the lesson and in turn motivate them to persevere through the end of the course.

Introduction

Computer Science is a growing field with many disciplines taught at many major colleges and universities in the United States. One of the foundational classes taught in a computer science major at universities according to the Association for Computing Machinery [ACM] is an Algorithms course. The purpose of this paper is to discuss why classrooms lack engagement in Computer Science and the ways to combat the lack of engagement of students. An implementation of how to keep a student engaged in a Computer Science course is explored with supplementary material in the form of Jupyter notebooks for a Udacity course called Introduction to Algorithms taught by Professor Michael Littman.

**Research Design**

A foundational Computer Science topic was chosen as introductory Computer Science classes typically see very high dropout rates of students [11]. According to ACM an algorithms course is apart of an official computer science curriculum because real-world performance of software systems depends on the algorithms chosen; therefore, good algorithm design is crucial for the performance of all software systems. Moreover, the study of algorithms provides insight into the intrinsic nature of the problem as well as possible solution techniques independent of programming language, programming paradigm, computer hardware, or any other implementation aspect [1].

I choose Professor Michael Littman's Introduction to Algorithms class on Udacity because essentially I wanted to choose a course that had the least amount of barriers to entry in education. An online course that is free and can be accessed by any device that has access to the internet seemed the most appropriate. The only requirement for Udacity is to sign up for a free account and the requirement for the course itself is an understanding of programming at the level of CS101, including the ability to read and write short programs in Python and a comfort level with mathematical notation at the high school Algebra II or SATs [2].

**1. Why Classrooms Lack Engagement**

1.1 Student Attention Spans

This day in age since the emergence of the internet in the late 1990s there has been a paradigm shift in the way people consume information. We are now accustomed to the immediate access to information by a quick internet search or even asking artificial intelligent assistants that come preloaded on cellphones, laptops, tablets, and even home appliances [7].

The habitual use of consuming information immediately but also in various forms has ingrained multi-stimuli and multitasking into young people's lifestyles. The lesson being taught to the student is therefore competing with the attention spans of these students who are not used to the slow traditional classroom setting of consuming information. [8].

1.2 Students Can't Relate to the Material

The younger generation is very education oriented with the educational pressure usually settling in around the first year of high school. The higher education system has only so many spots for future students that the bar to acceptance is constantly shifting ever higher where students need to separate themselves apart from the rest of the applicant pool. In fact, students often state their efforts in high school are a direct reflection of the type of college to which they plan to gain admission These students often complain that a particular subject seems unnecessary but not because they are expressing a lack of interest; instead the range of activities demanding their time and attention may make them less patient with lessons that do not directly apply to their goals or motivations in their chosen career path [8].

1.3 Instruction not Accommodating Students Learning Style

Research shows that how the brain works for individuals has added to the importance of developing and delivering curriculum and instruction that can match different learning styles of students. Not all students are the same therefore not all students learn the same way. Having a curriculum that can be flexible as far as communicating the same concept to a variety of learning needs is best for the overall comprehension of a class of students.

In order for a student to learn an appropriate level of challenge must match the student's' level or readiness to handle such difficulty. It is documented that if a student is engaged in a curriculum that is well beyond their ability; stressors result that impede learning for the student [9].

**2. Engagement Methodologies**

2.1 Chunking and Scaffolding

The ability to take a lesson and present it in a way for the student or students to understand can be challenging for an instructor. The student cannot be expected to immediately be on the same caliber of the instructor's understanding of the topic at hand who is presumably an expert in the field. The instructor needs to take the lesson being taught and separate it into easily digestible sections of concepts otherwise known as chunking [10].

For example, in an algorithms course one of the algorithms an instructor might present is the idea of sorting numbers least to greatest in value. Before the student can successfully implement an algorithm of this type they need to understand different concepts such as data types, functions, and reading output from the compiler of the programming language. These chunks of concepts must be mastered first before a more advanced concept such as the sorting algorithm can be understood.

Once the lesson is chunked into understandable parts the different concepts must be combined together though the teaching methodology of scaffolding. Scaffolding is a way for the teacher to provide support in the basic concepts and slowly introduce the student to more challenging problems [10]. The idea is that the student can take control of a more advanced application of the same problem and no longer need teacher support in the basic concepts contained in that same problem.

Following the same example as previously stated the goal was that the student can implement different types of sorting algorithms without the assistance of the teacher but now the learning activity the teacher presents is to choose which algorithm is more efficient given the background of the situation. Now the learning activity being employed is the concept of space and time when it comes to using sorting algorithms. This concept would have been completely loss on the student if it was presented before the basic application of how a sorting algorithm works.

## 2.2 Differentiation

A classroom of students can be challenging for both the individual student and the teacher. A teacher who is able to communicate a concept to different learning styles takes an understanding of each student to know their ability to handle the concept.

Curriculum needs to be differentiated to accommodate the learning abilities and learning styles of different students. Differentiation can be in the form of less challenging problem sets for the student who needs the extra information to solidify the concept into their current knowledge base to more challenging problem sets for the student who is feeling bored and is slowly lacking motivation or not feeling the content is not engaging enough for them to continue with the learning activity.

## 2.3 Bloom's Taxonomy

A framework was created by educational Psychologist Dr. Benjamin Bloom in order to help promote higher forms of thinking in education to help students move away from just remembering facts (rote learning) and is useful when designing educational, training, and learning processes [6]. There are six cognitive domains: remembering, understanding, applying, analyzing, evaluating, and creating. Remembering is the rote learning where one can recall a fact. Understanding is the ability to comprehend a subject. Applying is the use of the concept in a new situation. Analyzing is the ability to distinguish between facts and inferences. Evaluating is making judgements about the concept. Finally, creating is the ability to parts together to create a whole in a new situation. Bloom's Taxonomy can be useful in creating more engaging material because the learning activity will be utilizing different forms of thinking for the student.

## 3. Supplementary Content for Introduction to Algorithms

An overview of the Udacity course, the tools and technologies of the content is explained, and how the framework for how the notebooks are structured will be covered. A practical example of incorporating engagement strategies into the Udacity course is shown in the upcoming section

### 3.1 Introduction to Algorithms

Michael Littman's course on Udacity is well put together. The duration of the videos covers from about half a minute up to less than six minutes per video with most of the videos being a couple minutes long. As an experiment of participating in the course I felt like I was progressing at a comfortable rate. He is able to succinctly chunk a topic into logical and easy to understand parts. Notes to optional reading is provided for some videos for a deeper dive by the student. Engagement by students is seen in the forum for the course as students can ask and answer questions. Finally, there are quizzes as a check on learning throughout the course.

### 3.2 The need for the supplementary material

The need for the supplementary material as an addition to the class was explored. The course is well put together as Professor Littman is great at covering the introductory concepts and slowly move onto ideas that build on the previous lesson. The Jupyter notebooks were created for the student who needs the extra material to solidify the ideas being taught. One of the engagement strategies being implemented is differentiation. Some students may be able to cruise through the course alone while others may need various ways to have the concepts shown to them. The purpose of the notebooks is for the students who are experiencing self-efficacy and to keep them engaged in the class.

### 3.3 Tools and Technologies

The series of materials created for the class is built using Jupyter Notebooks. A Jupyter Notebooks is an open-source web application that allows you to create and share documents that contain live code, equations, visualizations and narrative text [3]. The notebooks give the educator the flexibility to communicate lessons in the form of various programming languages. The classroom exercises and notebooks were written in Python. The notebooks are available for consumption on Github [4].

### 3.4. The framework

The framework for the Jupyter Notebooks will be following the work of Ashlee Deline's paper "Exploring Engagement Through Supplementary Content for Linear Algebra" that was presented as part of the Summer 2017 Educational Technology class at Georgia Institute of Technology [5]. Each Jupyter Notebook contains six sections. Each section is one of the six cognitive domains of Bloom's Taxonomy which are in the following order: remembering, understanding, applying, analyzing, evaluating, and creating. One lesson in the class will have one notebook it is associated with and will be utilizing the chunking and scaffolding techniques to break down the lesson into concepts starting from basic level and then combined together to the advanced concepts at the end of the notebook.

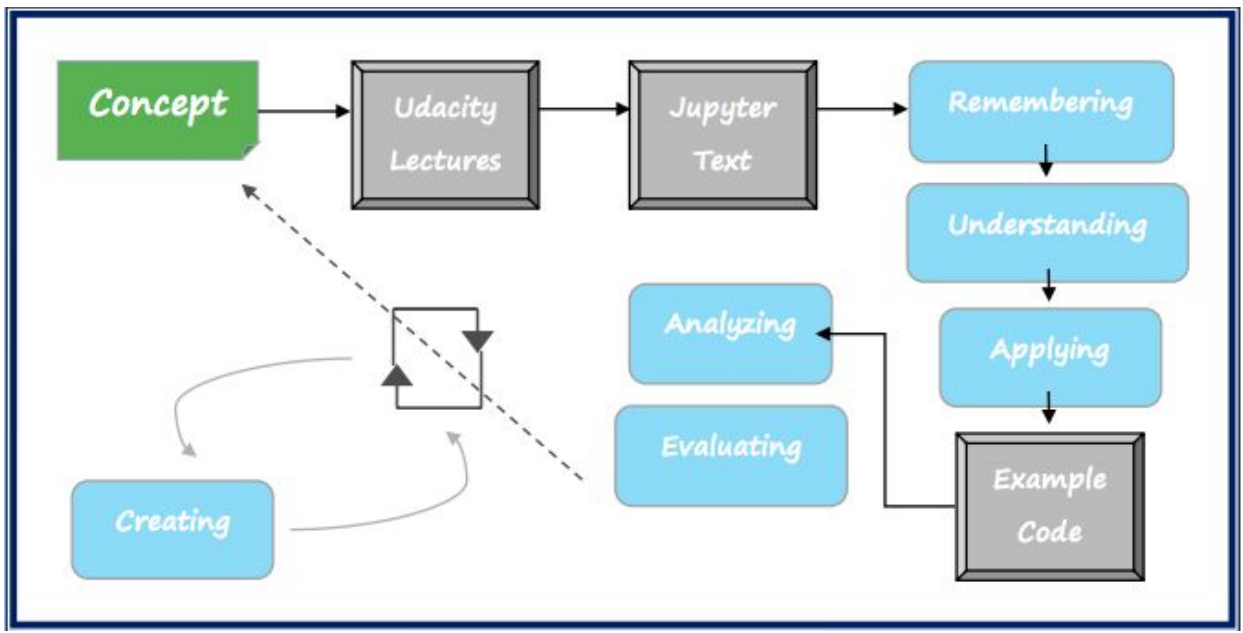

Figure 1: Jupyter Notebook Framework

**3.5 Class Overview**

The class outline for Introduction to Algorithms is below. There are seven lessons with the last section being interviews with various professionals in the field of Computer Science. Since the last lesson is not presenting any new material the decision was made to only provide Jupyter Notebooks for the first six lessons as supplementary material for the class.

- Introduction to Algorithms
    - Lesson 1: A Social Network Magic Trick
        - Become familiar with algorithm analysis
        - Eulerian Path ad Correctness of Na
        - Russian peasants algorithm and more
    - Lesson 2: Growth Rates in Social Networks
        - Use mathematical tools to analyze how things are connected.
        - Chain, ring and grid networks
        - Big Theta and more
    - Lesson 3: Basic Graph Algorithms
        - Find the quickest route to Kevin Bacon
        - Properties of social networks
        - Clustering coefficient and more
    - Lesson 4: It's Who You Know
        - Learn to keep track of your best friends using heaps
        - Degree centrality
        - Top K Via Partitioning and more
    - Lesson 5: Strong and Weak Bonds
        - Work with social networks that have edge weights
        - Make a tree and strength of connections
        - Weighted social networks and more
    - Lesson 6: Hardness of Network Problems
        - Explore what it means for a social network problem to be "harder" than other
        - Tetristan and Exponential Running Time

● Degrees of hardness and more
      ■ Lesson 7: Review and Application
            ● Interview with Peter Winker (Professor, Dartmouth College) on names and boxes problem and puzzles and algorithms
            ● Interview with Tina Eliassi-Rad (Professor, Rutgers University) on statistical measures in network and social networks in security and protests
            ● Additional interviews with Andrew Goldberg (Microsoft Research), Vukosi Marivate (Rutgers University) and Duncan Watts (Microsoft)

**3.6 Example content**

A Jupyter Notebook is made up of cells. The student will start at the top and run each cell. When a student is asked to answer a question they select their answer and run the cell which provides them feedback through colorful markings and audio confirmation. An explanation of each section is below.

*3.6.1 What You Will Learn*
The first section of each notebook is a brief overview of what they will learn after walking through the notebook's material. This section can be viewed as the goals they will accomplish before moving onto the next notebook. As shown in Figure 2 below the section puts the student into the mindset of what is ahead.

**What You Will Learn**

- Become familiar with algorithm analysis

- Eulerian Path

- Correctness

- Russian peasants algorithm

- Measuring timeee

Figure 2: What You Will Learn

*3.6.2 Remembering*

This second section of each notebook is called "Remembering." This section is based on the first cognitive domain of Bloom's Taxonomy. The goal of this domain was the calling or retrieval of previously learned information. This level is the most basic of cognitive domains as it provides the basis for all "higher" cognitive activity. For this section, multiple choice questions were utilized as shown in Figure 3.

Figure 3: Remembering Notebook Example

3.6.3 Understanding
The third section of each notebook is called "Understanding." This section is based on the second cognitive domain of Bloom's Taxonomy. The goal of this domain was the comprehension of the meaning, translation, interpolation, and interpretation of instructions and problems. For this section, matching was utilized as shown in Figure 4.

Figure 4: Understanding Notebook Example

*3.6.4 Applying*

The fourth section of each notebook is called "Applying." This section is based on the third cognitive domain of Bloom's Taxonomy. The goal of this domain was the use of a concept in a new situation or unprompted use of an abstraction. The student applies what was learned in the classroom into novel situations in the workplace. For this section, the student would create one of the algorithms in the course by hand as shown in Figure 5.

```python
# Eulerian tour
#
# Write a function create_tour that takes as input
# a list of nodes and outputs a graph as a list of tuples
# representing edges between nodes that have an Eulerian Tour.

def edge(x, y):
    return (x, y) if x < y else (y, x)

def create_tour(nodes):

    # Your code here

    return tour
```

Figure 5: Applying Notebook Example

*3.6.5 Analyzing*

The fifth section of each notebook is called "Analyzing." This section is based on the fourth cognitive domain of Bloom's Taxonomy. The goal of this domain was to separate material or concepts into component parts so that its organizational structure may be understood. The student is able to distinguish the difference between facts and inferences. For this section, the student undergoes multiple choice, multiple answer questions as shown in Figure 6.

Figure 6: Analyzing Notebook Example

### 3.6.6 Evaluating

The sixth section of each notebook is called "Evaluating." This section is based on the fifth cognitive domain of Bloom's Taxonomy. The goal of this domain was to make judgements about the value of ideas or materials. For this section, the student evaluates if the statement made is true or false and if false explain as shown in Figure 7.

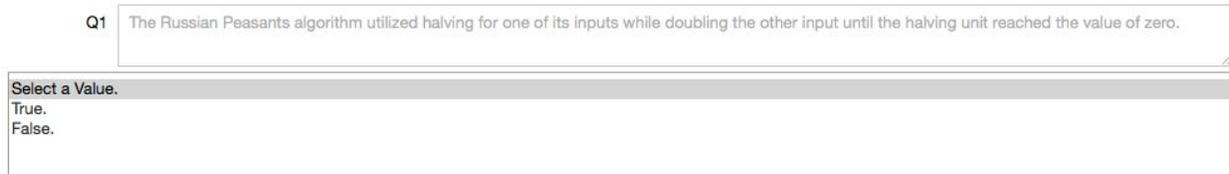

Figure 7: Evaluating Notebook Example

### 3.6.7 Creating

The seventh section of each notebook is called "Creating." This section is based on the sixth and final cognitive domain of Bloom's Taxonomy. The goal of this domain was to build a structure or pattern from diverse elements. The student is able to put parts together to form a whole, with emphasis on creating a new meaning or structure. For this section, the student is asked to go beyond the lesson by applying previously learned chunks of knowledge to create something new as shown in Figure 8.

```python
# In the lessons we learned about the naive algorithm

def naive(a, b):
    x=a; y=b
    z=0
    while x > 0:
        z = z + y
        x = x - 1
    return x

# Also in the lessons we learned about recursion - a method where the function calls itself
# Implement a recursive version of the naive algorithm below:
```

Figure 8 Creating Notebook Example

## 3.6 Future Scope

There are many future enhancements that can be made to create more value for the student. The notebooks are available open source and hosted on Github so collaboration by other developers is possible. One of the many enhancements can be cosmedic to make the notebooks more appealing. The widgets can be refactored for aesthetics and functionality. Adding more content for the student would make the notebooks more useful. The backend code can be cleaned up to implement best practices such as "don't repeat yourself", making the code idiomatic, and more code documentation to make the code easier to understand.

**4 Conclusion**

Engagement of the student in the early stages of Computer Science field is critical to their success as the progress to more advanced concepts that are built on top of foundational ideas. One of the foundational topics in Computer Science that was explored is Algorithms and supplementary materials were created for the Introduction to Algorithms by Professor Michael Littman on Udacity. The notebooks were based on a framework created by Ashlee Deline of Georgia Institute of Technology, Educational Technology Summer 2017 course. The goal of the project was to create engaging material to help students experience self-efficacy and to motivate them to keep being engaged in the class.

**Acknowledgements**

**References**

[1] Curriculum Guidelines for Undergraduate Degree Programs in Computer Science. (2017, December 20). Retrieved December 7, 2017, from https://www.acm.org/binaries/content/assets/education/cs2013_web_final.pdf
The Joint Task Force on Computing Curricula Association for Computing Machinery (ACM) IEEE Computer Society

[2] Introduction to Algorithms | Udacity. (n.d.). Retrieved Dec 7, 2017, from https://www.udacity.com/course/intro-to-algorithms--cs215

[3] Project Jupyter. (n.d.). Retrieved May 27, 2017, from http://www.jupyter.org

[4]Birster, Chris, Intro-Algorithms, (2017), GitHub repository, https://github.com/chrisbirster/Intro-Algorithms

[5] Deline, A. (2017, July). Exploring Engagement Through Supplementary Content for Linear Algebra. Retrieved December 07, 2017, from t-square.gatech.edu

[6] Bloom's Taxonomy of Learning Domains. Knowledge Jump. Clark, Donald. June 5, 1999. Retrieved May 25, 2017, from http://www.nwlink.com/~donclark/hrd/bloom.html

[7] Tang, Bingqing, "The Emergence of Artificial Intelligence in the Home: Products, Services, and Broader Developments of Consumer Oriented AI" (2017). *Student Theses, Papers and Projects (Computer Science)*. 6. https://digitalcommons.wou.edu/computerscience_studentpubs/6

[8] Barnes, K., Marateo, R. C., & Ferris, S. P. (2007, May). Innovate: Journal of Online Education. Retrieved December 07, 2017, from http://nsuworks.nova.edu/innovate/vol3/iss4/1/

[9] Tomlinson, C. A., & Kalbfleisch, M. L. (1998). Teach Me, Teach My Brain: A Call for Differentiated Classrooms. *Educational Leadership*, *56*(3), 52-55.

[10] R. Hansen, S & H. Narayanan, N. (2000). Helping learners visualize and comprehend algorithms. Interactive Multimedia Electronic Journal of Computer-Enhanced Learning. 2.

[11] Aharon Yadin, Reducing the dropout rate in an introductory programming course, ACM Inroads, v.2 n.4, December 2011 [doi>10.1145/2038876.2038894]

[12] ipywidgets — ipywidgets and jupyter-js-widgets 7.0.0a3 documentation. (n.d.). Retrieved Dec 7, 2017, from https://ipywidgets.readthedocs.io/en/latest/

[13] Matplotlib. Retrieved Dec 7, 2017, from Matplotlib, https://matplotlib.org/

[14] NumPy. Retrieved Dec 7, 2017, from http://www.numpy.org/

[15] Using Audio in IPython. Retrieved Dec 7, 2017, from http://musicinformationretrieval.com/ipython_audio.html